\documentstyle[12pt,epsf,epsfig,wrapfig]{article}
\begin{document}
\begin{center}
{\bfseries SPIN ALIGNMENT OF VECTOR MESONS IN POLARIZED AND 
UNPOLARIZED HIGH ENERGY REACTIONS}

\vskip 5mm
Liang Zuo-tang
\vskip 3mm

{\small
Department of Physics, Shandong University, Jinan, Shandong 250100, CHINA}

\end{center}

\vskip 5mm

\begin{abstract}
Recent $e^+e^-$ annihilation experiments show a striking spin alignment 
of vector mesons produced in the fragmentation of a polarized quark.
We show that these data imply a simple relation between the polarization 
of the anti-quark, which is produced in the fragmentation process to 
combine with the fragmenting quark to form the meson, and that of the 
fragmenting quark. After having obtained a good fit to the $e^+e^-$ data, 
we extend the calculations to lepton-nucleon and nucleon-nucleon 
collisions with polarized beams. We also discuss the relation 
between the spin alignment of vector meson in unpolarized hadron-hadron
collisions and the single-spin left-right asymmetry.  We make predictions 
for the polarization of the vector mesons in these reactions which can be 
checked by future experiments.
\end{abstract}

\vskip 8mm

\section{Introduction}
Polarization experiments have been proven to be a powerful tool 
to study the properties of strong interaction in general and to 
test the validity of popular models in particular. 
Great achievements have been made in the last years 
by studying the spin-dependent structure functions, 
the single-spin left-right asymmetries, 
hyperon polarizations etc, both experimentally and theoretically. 
In this talk, I would like to bring your attention to 
another series of polarization experiments 
which may also provide useful information on strong interaction.

The experiments which I mention are the measurements of 
vector meson polarization in high energy reaction. 
I emphasize here in particular the following:
(1) Vector meson polarization can be studied experimentally 
by measuring the angular distribution of the decay products.
(2) Compared with other spin non-zero hadrons such as the hyperons,
the production rate of a vector meson is usually higher
and the origin is simpler
due to the much smaller contribution from the decay of heavier hadrons.
This implies less theoretical uncertainties in calculations and
better statistics in experiments. 
Therefore, the study of polarization of the vector mesons may provide us
important information to understand the role of spin
in hadronic reactions.

The polarization of a vector meson is
described by the spin density matrix $\rho$ or its element $\rho_{m,m'}$,
where $m$ and $m'$ label the spin component along the quantization axis,
usually the $z$-axis of the frame.
The diagonal elements $\rho_{11}$, $\rho_{00}$ and $\rho_{-1-1}$
for the unit-trace matrix are the relative intensities of meson spin
component $m$ to take the values $1$, $0$, and $-1$ respectively.
In experiment,
$\rho_{00}$ can be measured from the angular
distribution of its decay products and
a deviation of $\rho_{00}$ from $1/3$ indicates spin alignment.
In the helicity basis, i.e., in the case that the
$z$-axis is chosen as the moving
direction of the vector meson, the matrix is usually called
helicity density matrix and
$\rho_{00}$ represents the probability
of the vector meson to be in the helicity zero state.
Measurements have been carried out in different
reactions\cite{Savhh,epr82,BCGMNS,BEBC87,EXCHARM00,LEP}. 
We see in particular that data with high accuracy 
have been obtained in $e^+e^-$ annihilation at LEP recently \cite{LEP}.
The data\cite{LEP} show that
$\rho_{00}$ is significantly larger than $1/3$ in particular 
in the large $z$ (here $z$ is the fractional momentum carried by 
the vector meson of the whole system) region. 

Being not aware of what we can learn from these data, we made an analysis 
by taking the spins of the vector mesons as the sum of the spin 
of the fragmenting quark and that of the anti-quark which is created 
in the hadronization and combine with the fragmenting quark to form 
the vector meson. 
We found out that the data imply a significant polarization of  
the anti-quark in the opposite direction of the fragmenting quark. 
We extended the calculations to other reactions to make predictions 
for future experiments. 
We also studied the situation in unpolarized hadron-hadron collisions 
and showed that the spin alignment of vector meson in such reactions 
should be closely related to the single-spin left-right asymmetries. 
I will summarize these results in the following. 
The material of this talk is taken from the recent publications\cite{XLL01,XL02} 
carried out in collaboration with Dr. Xu Qing-hua and Dr. Liu Chun-xiu 
in Shandong University.

\section{Spin alignment of vector mesons in $e^+e^-\to VX$}
We calculate the spin alignment of vector mesons in 
$e^+e^-\to q_f^0\bar q_f^0\to VX$ by 
dividing them into the following two groups
and considering them separately:
(A) those which contain the fragmenting quark $q^0_f$;
(B) those which don't contain the fragmenting quark.
The spin density matrix $\rho^V(z)$ for the vector meson $V$
is given by:          
\begin{equation}
\rho^V(z)=
\sum_f \frac {\langle n(z|A,f)\rangle}{\langle n(z)\rangle}\rho^V(A,f)
+\frac {\langle n(z|B)\rangle}{\langle n(z)\rangle}\rho^V(B),
\label{eq1}
\end{equation}
where $\langle n(z|A,f)\rangle$ and $\rho^V(A,f)$
are respectively the average number and the spin density matrix 
of the vector mesons from group (A);
$\langle n(z|B)\rangle$ and $\rho^V(B)$ are those from (B);
$\langle n(z)\rangle=\sum_f \langle n(z|A,f)\rangle+
\langle n(z|B)\rangle$, 
and $z$ is the momentum fraction of $q_f^0$ 
carried off by the vector meson $V$.
Since $\langle n(z|A,f)\rangle$
and $\langle n(z|B)\rangle$ are independent of the spin properties
they are calculated using an
event generator based on a fragmentation model 
which gives a good description of the unpolarized data.
We used generator $\sc pythia$ in our calculations.

There are many different possibilities to produce 
the vector mesons in group (B), 
we take them as unpolarized, i.e., $\rho^V(B)$=1/3.
For those vector mesons from group (A),
i.e., those which contain $q^0_f$ and
an anti-quark $\bar q$ created in the fragmentation,
the spin is taken as the sum of the spins of
$q_f^0$ and $\bar q$,
thus the spin density matrix $\rho^V(A,f)$ can be calculated
from the direct product of the spin density matrix
$\rho^{q^0_f}$ for $q^0_f$
and  $\rho^{\bar{q}}$ for $\bar{q}$.
Transforming the direct product,
$\rho^{q^0_f{\bar q}}$=$\rho^{q^0_f}$$\otimes$$\rho^{\bar q}$,
to the coupled basis $|s, s_z\rangle$ 
(where $\vec{s}$=${\vec {s}^q}$+${\vec {s}^{\bar q}}$),
we can obtain $\rho^V(A,f)$.
In the helicity frame of $q_f^0$, i.e., $z$-axis is taken as
the moving direction of $q_f^0$, $\rho^V(A,f)$ is given by, 
\begin{equation}
\rho^V(A,f)=\frac{1}{3+P_fP_z}
\small {
\left (
\begin{array}{ccc}
(1+P_f)(1+P_z) & \frac{{1+P_f}}{\sqrt2}(P_x-iP_y)   & 0      \\
\frac{(1+P_f)}{2}(P_x+iP_y) &  (1-P_fP_z)&  \frac{{1-P_f}}{\sqrt2}(P_x-iP_y) \\
0 & \frac{1-P_f}{\sqrt2}(P_x+iP_y) &  (1-P_f)(1-P_z)          \\
\end{array}  \right), }
\label{eq2}
\end{equation}            

\begin{wrapfigure}{R}{8cm}
\mbox{\epsfig{figure=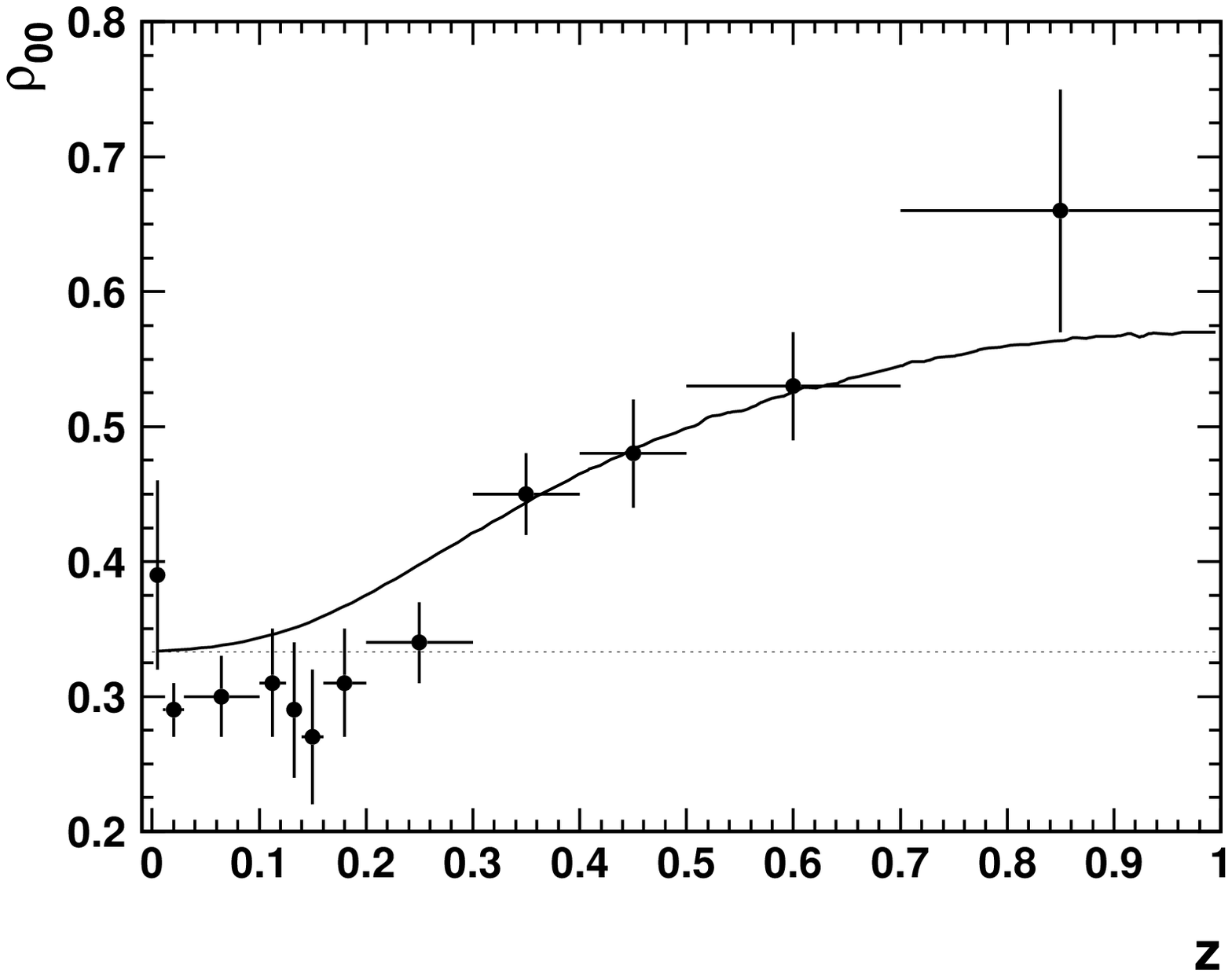,width=7.6cm}}
\vskip -0.5cm
{\small{\bf Figure 1.} Spin alignment of $K^*$ in $e^+e^-$ annihilation at $Z$ pole.
The figure is taken from [7].}
\medskip
\end{wrapfigure}

\noindent
where $P_f$ is the longitudinal polarizations of $q^0_f$ and
$\vec P=(P_x,P_y,P_z)$ is the polarization vector of 
the anti-quark $\bar q$.
Hence, the 00-component of the density matrix takes 
the following simple form,
\begin{equation}
\rho_{00}^V(A,f)=(1-P_fP_z)/ (3+P_fP_z),
\label{eq3}
\end{equation}                   
where there is only one unknown variable $P_z$, 
the polarization of $\bar q$ along the 
moving or polarization direction of $q_f^0$.

Using Eqs.(\ref{eq1}) and (\ref{eq3}), we can
determine $P_z$ in different cases by fitting 
the data\cite{LEP} in the $e^+e^-$ annihilation at $Z^0$ pole 
for the production of different vector mesons.
We found that\cite{XLL01},
the data for different vector mesons can reasonably be fitted 
if we take,
\begin{equation}
 P_z=-\alpha P_f
\end{equation} 
where $\alpha \approx 0.5$. 
As an example, we show $\rho_{00}^{K^*}$ as 
a function of $z$ in Fig.1.

\section{Spin alignment of vector mesons in $pp\to XV$ 
and $lp\to l'VX$ with polarized beams}

We emphasize that the relation between $P_z$ and $P_f$ 
given in Eq.(4) is a consequence of $e^+e^-$ data at LEP. 
It is unknown whether it is universal in the sense that 
it is also applicable to other energies and/or other reactions. 
If yes, it should contain useful information on the hadronization mechanism. 
We therefore extended the calculations to other reactions 
such as $pp\to VX$ or $lp\to l'VX$ with polarized beams in [8]. 
The results obtained show that the spin alignments of vector mesons 
in these reactions are all quite significant and should be able to tested 
by future experiments such as those at RHIC and/or by COMPASS.
I show, as examples, the results for $pp\to VX$ and 
$\mu^-p\to\mu^-VX$ in Fig.2 and Fig.3.

\begin{tabular}{ll}
\begin{minipage}{7cm}
\mbox{\epsfig{figure=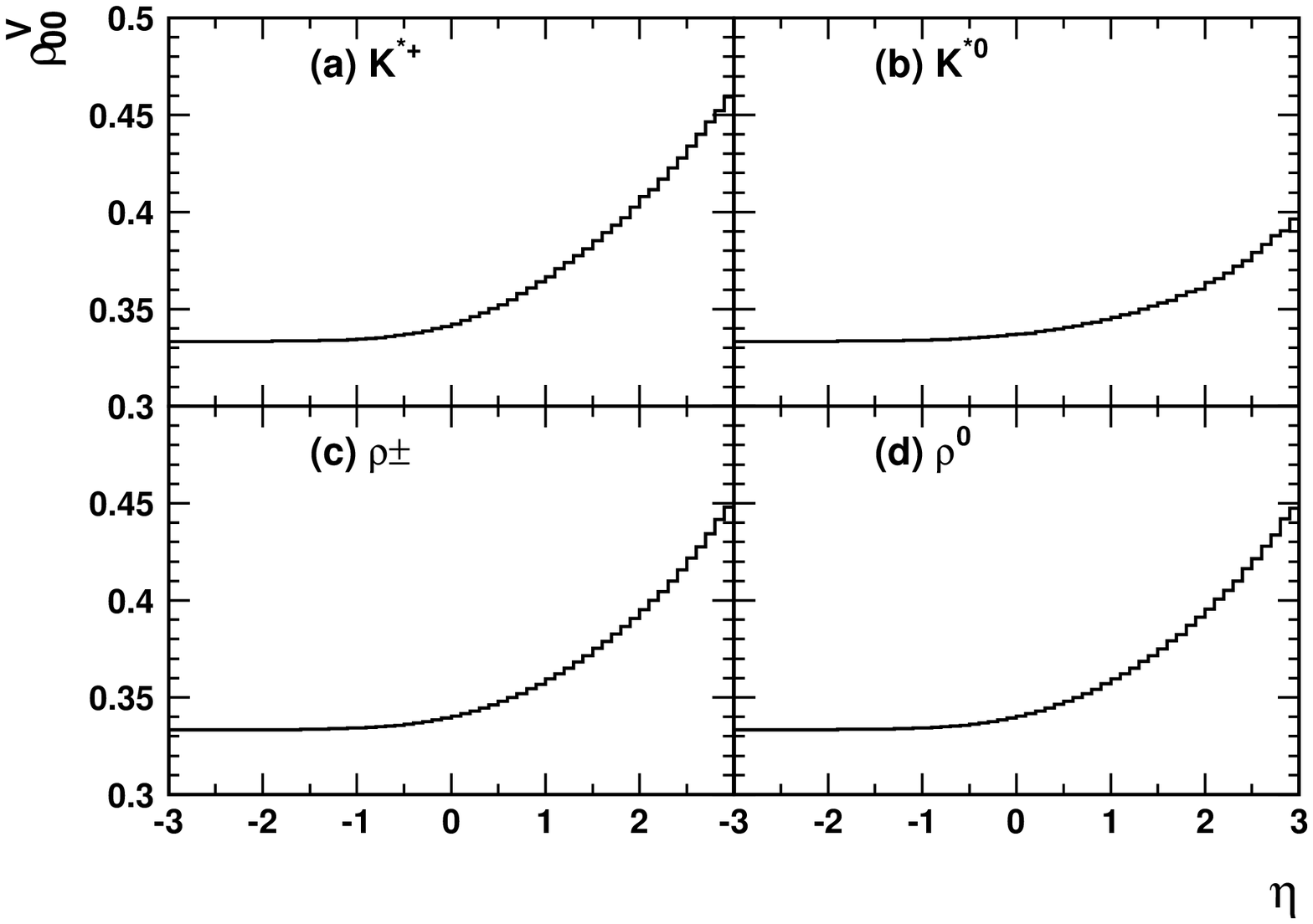,width=7cm}}
\vskip -0.5cm
{\small{\bf Figure 2.} $\rho^V_{00}$ 
in $pp\to VX$ for $p_\perp>13$ GeV at $\sqrt{s}=500$GeV 
in the helicity frame
when one beam is longitudinally polarized.
The figure is taken from [8].}
\end{minipage} &
\begin{minipage}{7cm}
\mbox{\epsfig{figure=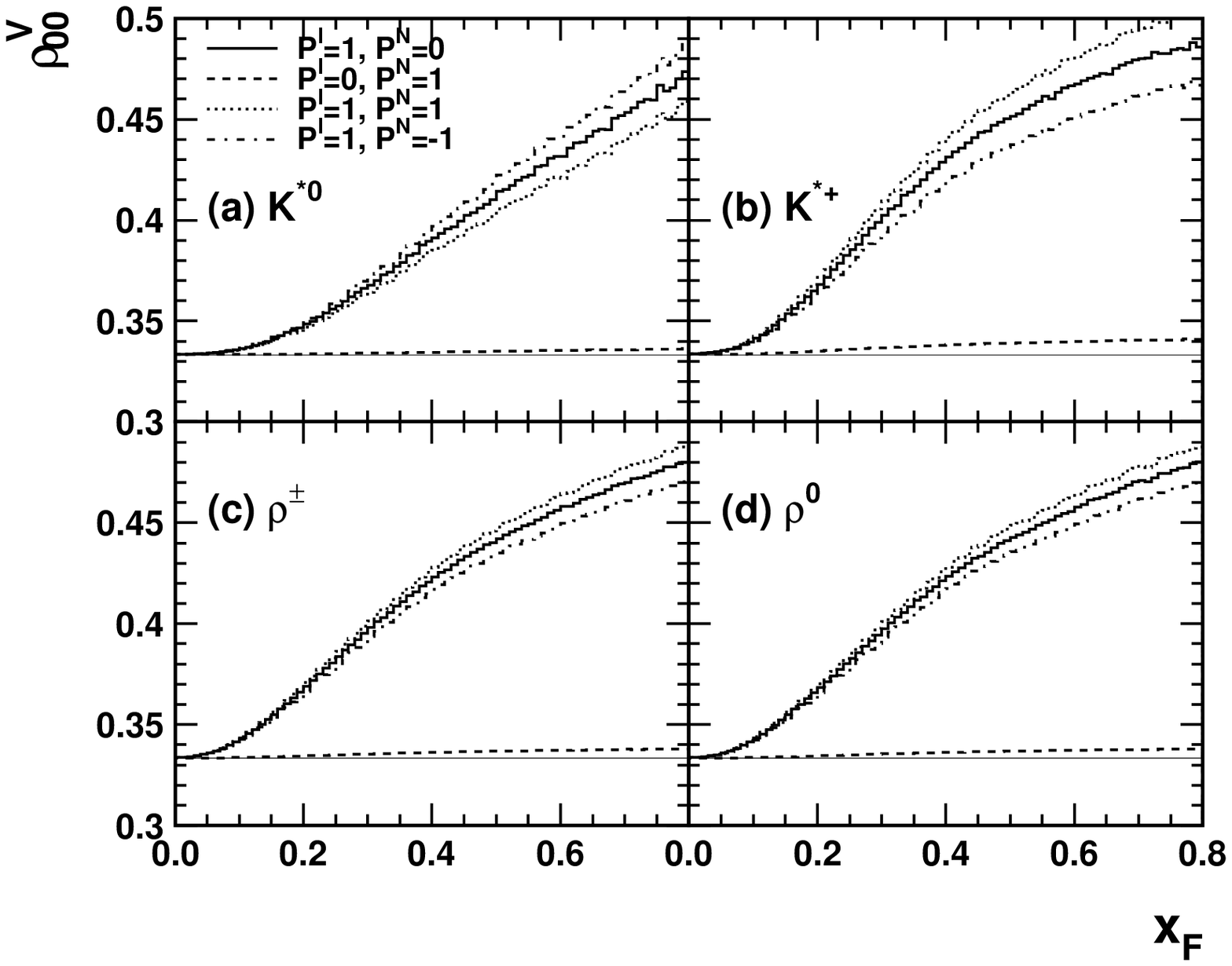,width=7cm}}
\vskip -0.5cm
{\small{\bf Figure 3.}$\rho^V_{00}$ in the current
 region of $\mu^- p$$ \to$$ \mu^- VX$ at $E_\mu$=500 GeV.
The figure is taken from [8].}
\end{minipage}\\
\end{tabular}

\section{Spin alignment of vector mesons in unpolarized hadron-hadron collisions}

Another interesting phenomenon in this connection 
is the spin alignment 
of vector meson in unpolarized hadron-hadron collision.
The situation here is quite similar to transverse 
hyperon polarization in unpolarized hadron-hadron collision.
The quantization axis is chosen as the normal of the production plane 
and the data show that there is a significant spin alignment 
of vector meson at large $x_F$ and moderate $p_\perp$.

\begin{wrapfigure}{R}{8cm}
\mbox{\epsfig{figure=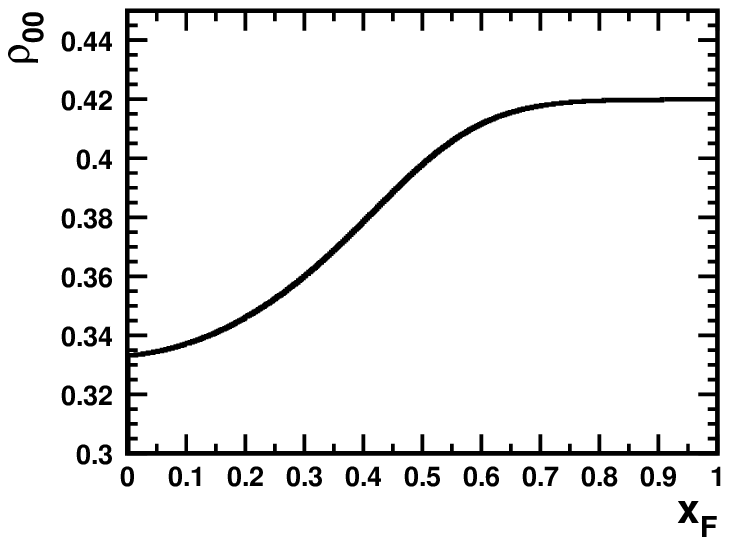,width=7.6cm}}
\vskip -0.3cm
{\small{\bf Figure 4.} Spin alignment of $K^{*+}$ along the normal of the
production plane in unpolarized $pp\to K^{*+}X$ at $p_{inc}=200$ GeV.
The figure is taken from [8].}
\medskip
\end{wrapfigure}

We argued that the existence of the three classes of phenomena,
 i.e. the large left-right asymmetries $A_N$ in hadron-hadron collisions 
with transversely polarized beams\cite{ANdata}, 
the large transverse polarization of hyperon $P_H$ 
in unpolarized $hh$ or $hA$ collisions\cite{PHdata},
and significant spin alignment of vector mesons 
$\rho_{00}^V$ in unpolarized $hh$ collisions,
are different manifestations of the same spin correlations of the type
$\vec s_q \cdot \vec n$ in the reaction.
[Here, $\vec s_q$ is the spin of the fragmenting quark;
$\vec n\equiv (\vec p_{inc}\times \vec p_h)/|\vec p_{inc}\times \vec p_h|$ 
is the unit vector in the normal direction of the production plane, 
$\vec p_{inc}$ and $\vec p_h$ are respectively 
the momentum of the incident
hadron and that of the produced hadron.] 
We demonstrated the close relation between $A_N$ 
and $P_H$, and that between $A_N$ and $\rho_{00}^V$ 
in unpolarized hadron-hadron collisions 
in the following way: 
We note that the three effects exist mainly in the fragmentation 
regions and hadrons produced in these regions are mainly 
products of the valence quarks combined with suitable anti-sea-quark 
or sea-diquarks. 
The existence of spin correlation $\vec s_q \cdot \vec n$
implies that the valence quark which combines with suitable anti-sea-quark 
(or quarks) to form the produced hadron was transversely polarized to 
the production plan. 
We use the results derived from the $A_N$ data\cite{ANdata} 
for the strength of the above-mentioned $\vec s_q\cdot \vec n$ 
type of spin-correlation as input to 
calculate the polarization of the valence-quark 
before the hadronization. 
Using the result obtained as input, we 
calculated the hyperon polarization in \cite{LB97} 
and spin alignment of the vector mesons in \cite{XL02}. 
The obtained results are consistent with the available data 
and predictions for future experiments are made. 

For vector meson spin alignment in unpolarized $hh$ collisions, 
the results obtained are quite different for 
the vector mesons of the following two groups 
according to their flavor compositions: 
(1) those which have a valence quark of
the same flavor as one of the valence quarks of the incident hadron; 
(2) those which have no valence quark 
in common with the incident hadron.
The behavior of $\rho_{00}^V(x_F)$ 
as a function of $x_F$ for each group is similar 
but is quite different from each other. 
For those from group (1), 
$\rho_{00}^V(x_F)$ increases with increasing $x_F$.
It starts from $1/3$ at $x_F=0$, 
increases monotonically with increasing $x_F$, 
and reach about $0.42$ as $x_F\to 1$.
A rough estimation of 
the shape of the $x_F$-dependence of $\rho_{00}^V(x_F)$ 
is made and the results are given in Fig.4.
For those from group (2), 
we expect no significant spin alignment,  
i.e. $\rho_{00}^V(x_F)\approx 1/3$ for all $x_F$.
All these can be checked by future experiments.

I thank the organizer for invitation. This work was supported 
in part by the National Science Foundation of China (NSFC) 
and Education Ministry of China.


\end{document}